\documentclass[prl,groupedaddress,twocolumn,showpacs]{revtex4-1} 
\usepackage{epsfig}
\usepackage{dcolumn}
\usepackage{epstopdf}
\usepackage{graphicx}
\usepackage{ulem}
\usepackage{color}
\begin{document}
\newcommand{\new}[1]{\textcolor{blue}{#1}}
\newcommand{\old}[1]{\textcolor{red}{#1}}
\title{Ordering and multiple phase transitions in ultra-thin nickelate superlattices
}  
\author{Danilo Puggioni}\author{Alessio Filippetti} 
\author{Vincenzo Fiorentini}\affiliation{Dipartimento di Fisica dell'Universit\`a di Cagliari and CNR-IOM, UOS Cagliari,\\ Cittadella Universitaria, I-09042 Monserrato (CA), Italy}\date{\today}
\begin{abstract}
We  interpret via  advanced ab initio   calculations  the multiple  phase transitions observed recently   in ultra-thin LaNiO$_{3}$/LaAlO$_{3}$ superlattices. The ground state is  insulating, charge-ordered, and antiferromagnetic due to concurrent structural distortion and  weak valency disproportionation. We infer distinct transitions around 50 and 110 K, respectively, from antiferromagnetic order to moment disorder, and from structurally-dimerized insulator to an undistorted metallic Pauli paramagnet (exhibiting a cuprate-like Fermi surface). The results are in satisfactory agreement with  experiment. \end{abstract}
\pacs{71.30.+h,
75.25.Dk,
75.70.-i
}
\maketitle
\noindent

Incipient or actual  instabilities   towards collective ordered states are typical of many correlated materials.  Recent  experiments  \cite{boris,mottlno} have investigated,  via  elegant nanostructuring manipulations of materials properties, the phase transitions playing out when LaNiO$_{3}$ (LNO henceforth -- in the bulk the only 
metallic Pauli-paramagnetic (PM)   rare-earth nickelate) is placed in an intentionally perturbed environment, namely 
epitaxially-strained  ultra-thin superlattice (SL) of  alternating layers of LNO and of the band insulator LaAlO$_{3}$ (LAO).  For sufficiently thin LNO (2-3 layers at most),  multiple transitions from a non-magnetic normal metal to a long-range-ordered magnetic, insulating, charge-ordered state were revealed by a crossover in  conductivity temperature (T) dependence, muon spin rotation ($\mu$SR), spectral weight transfer  in  optical conductivity  \cite{boris}, and XAS (x-ray absorption spectroscopy) line splitting  \cite{mottlno}.  Magnetometry and  $\mu$SR strongly support   long-range antiferromagnetic (AF) order \cite{boris}.
 
 The  precise  nature of the low-T state of the LNO/LAO SL and the transitions it undergoes is unclear.  Here we  address the problem from first principles studying  a strained ultra-thin   LNO/LAO SL  using variational self-interaction-corrected local density functional theory (VPSIC) \cite{vpsic,ff}, a parameter-free method   providing an improved description of correlated and magnetic materials     compared to semi-local approaches. So far, ab initio   calculations  have neither been able to identify an AF ground state as observed  for these SLs, nor, a fortiori, to provide a general picture of the rich  experimental  situation. Local  (LDA) or gradient-corrected (GGA) density functionals  find neither a  stable AF phase nor distorted structures. 
 LDA/GGA+U  predicts ferromagnetic ground states not seen in experiment, apparently  irrespective of U \cite{Ulno}. The     satisfactory agreement of VPSIC   with  experiment demonstrated below suggests an improved  account for on-site correlations, also indicated by points of agreement with dynamical mean field theory (DMFT) \cite{notedmft,held,held2,millis}, especially with the interpretation proposed in \cite{millis} during the review of the present work.

 The ground state, labeled AFD henceforth,  is structurally dimerized,  weakly charge-ordered, insulating, and AF with in-plane modulation similar to bulk rare-earth nickelates \cite{nickmag}. From  calculated energies, we infer magnetic-ordering and metal-insulator transitions at two distinct critical temperatures of around 50 K  and 110-150 K respectively. 
The transitions are driven by cooperative  structural distortion and  partial valency disproportionation of Ni atoms at low T, inducing magnetic superexchange and Mott localization.  In addition, we calculate the T-dependent SL conductivity  within  Bloch-Boltzmann theory \cite{bbt} and discuss a possible  concurrent  metal-insulator-transition mechanism.
The results are  consistent with the   data  of Ref.\onlinecite{boris}. 
We find that the high-T metallic Pauli-PM phase has a  Fermi surface  geometrically (though not orbitally) akin to optimally-doped cuprates, as suggested  earlier  \cite{kha,held}. 

\paragraph{Method$\!\!\!\!$}-- Total energy, force, and bands calculations are performed by VPSIC \cite{vpsic} using the plane-wave ultrasoft pseudopotential method (energy cutoff 30 Ryd,   4$\times$4$\times$4 k-mesh, Gaussian smearing of 20 mRyd for the metal phases) in an 80-atom  2$\sqrt{2}$$\times$2$\sqrt{2}$$\times$2 perovskite supercell {(40-atom  2$\sqrt{2}$$\times$$\sqrt{2}$$\times$4 only for the calculation of vertical magnetic coupling)}.  The LNO/LAO (1+1) SL is simulated at the in-plane lattice constant 4.02 \AA, corresponding to a tensile planar strain of about 4\% relative to the LNO bulk lattice constant. The   LNO layer  contains four Ni atoms to simulate AF structures. We optimize the cell length, and atomic positions via  quantum forces  \cite{vpsic}. The {\it dc} conductivity is calculated in Bloch-Boltzmann  theory   \cite{bbt} with a relaxation time approximation
using ab initio band energies on dense k grids (over 2000 points).

\begin{figure}[h]
\includegraphics[width=7 cm]{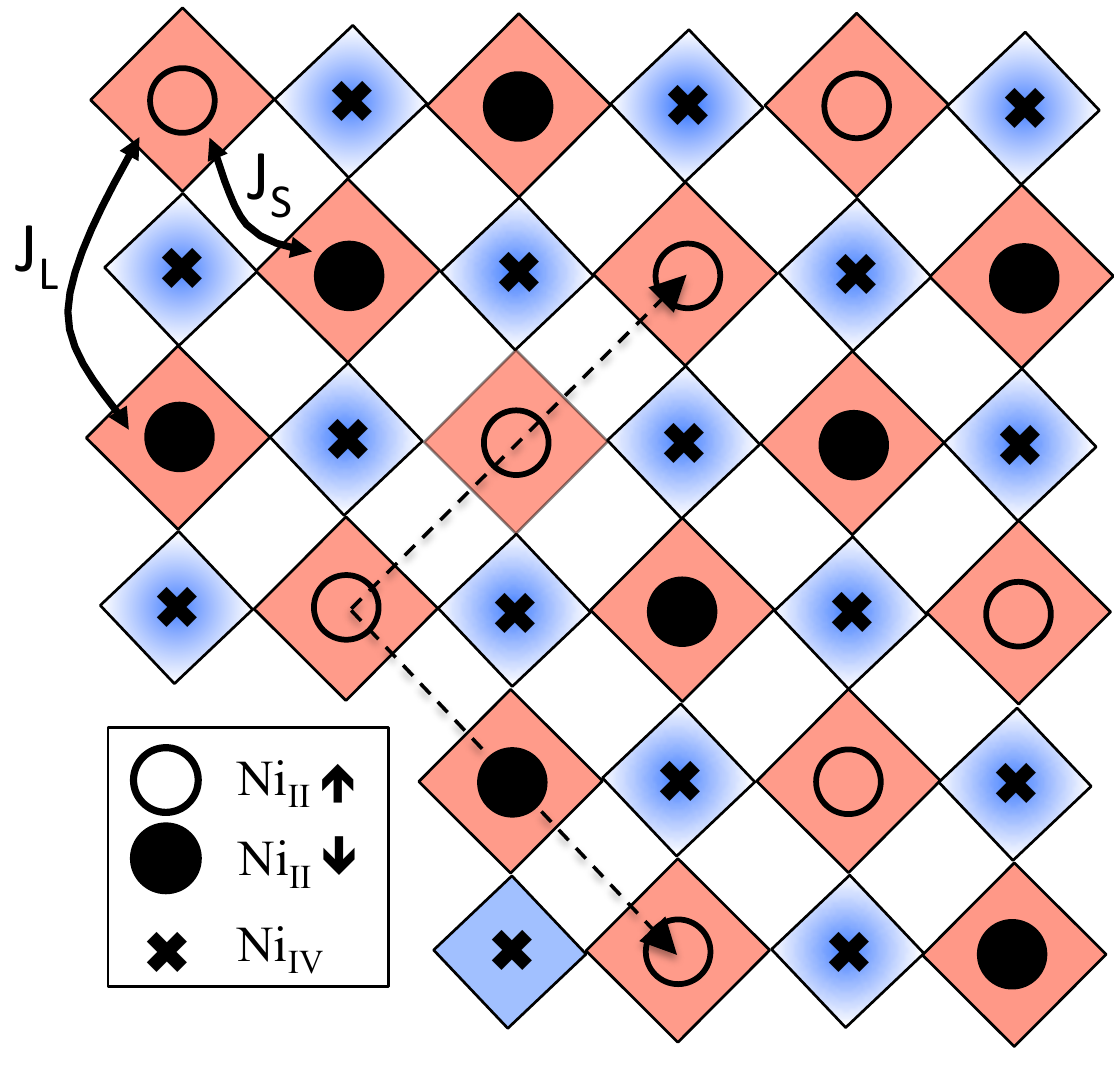}
\caption{(Color on line) Magnetic pattern of the LNO layer of LNO/LAO (1+1). 
Expanded  and contracted octahedra around Ni$_{\rm II}$ and Ni$_{\rm IV}$, respectively, are red (filled) and blue (shaded). Oxygens (not shown) sit at the shared vertexes of the octahedra. Dashed: supercell lattice vectors.  Drawing is approximately to scale.}
\label{figmag}
\end{figure}

\paragraph{Structure and charge ordering$\!\!\!\!$}--  
In the AFD ground state the nominally trivalent Ni$_{\rm III}$ atoms of LNO are  actually inequivalent in pairs due to a strong cooperative ``checkerboard'' dimerization of the Ni-centered octahedra (Fig.\ref{figmag}).  
Ni-O bonds  
alternatively expand and  contract from  2 \AA\, to 2.19 \AA\, or  respectively 1.83 \AA\, in  a predominantly breathing mode. Calculated octahedra rotations are  minor ($\sim$1-2$^{\circ}$).  The  short and long bonds  match those, respectively,   in peroxonickel complexes with nominally tetravalent Ni$_{\rm IV}$ and NiO with  nominally divalent Ni$_{\rm II}$. Although the charge transfer is small, we adopt this labeling convention for clarity.  The distortion is indeed accompanied by charge transfer from Ni$_{\rm IV}$ to Ni$_{\rm II}$, which we  quantify by VPSIC occupations  \cite{vpsic}. The total transfer is 0.07 $|e|$, in fair agreement with 0.03 $|e|$ estimated  in \cite{boris}.  
 This charge-ordered bond-dimerized phase  is quite compatible with the splitting in the SL XAS spectra, analogous to insulating nickelates \cite{mottlno}, although we cannot  provide a quantitative estimate of   that splitting from our calculation. The magnetic order and insulating character of this  state tend to confirm this conclusion, as discussed below. 

%
%
%
%
\paragraph{Magnetism$\!\!\!\!$}--  The  structural and valency dimerization are associated with the  magnetic pattern   in Fig.\ref{figmag}.  Our single-LNO-layer  cell  has   no vertical modulation by construction; the planar modulation {\bf q}=(0,1/2) in the reciprocal basis of the supercell (Fig.\ref{figmag}) is   the same as in bulk monoxides \cite{vpsic}, and it is analogous to insulating nickelates. The main difference is that Ni$_{\rm IV}$ is entirely non-magnetic here, while the homologous ``Ni(3--$\delta$)'' are still polarized in e.g. NdNiO$_{3}$ \cite{nickmag}. Indeed,  Ni$_{\rm II}$'s [circles in the Figure] carry a moment $\mu_{\rm Ni_{\rm II}}$=$\pm$1.44 $\mu_{B}$, while Ni$_{\rm IV}$'s [crosses in the Figure] have zero moment, confirming a qualitative picture of Ni$_{\rm III}$ disproportionation into unpolarized Ni$_{\rm IV}$ $t_{2g}^{6}e_{g}^{0}$ and polarized Ni$_{\rm IV}$ $t_{2g}^{6}e_{g}^{2}$. Overall, our charge-spin order pattern
%
    matches closely  the 
mechanism sketched in Fig.1b of Ref.\cite{mazin};  on-site exchange and structural energy gains overrule the effective on-site repulsion, quite screened due to the e$_g$ states delocalization  (which is, in turn, coherent with  LNO being paramagnetic, and with a weakened Jahn-Teller effect). Our result clearly agrees with the AF long-range order suggested by   magnetometry and $\mu$SR. 

The AFD phase is governed by the in-plane couplings $J_{L}$ and $J_{S}$ (Fig.\ref{figmag}), and vertical coupling  $J_{\perp}$ across the LAO layer.  An AF $J_L$ is expected due to  superexchange between partially-filled $e_g$ states. $J_{S}$ would  be AF for purely $x^{2}$--$y^{2}$ $e_{g}$ hopping, but as  $e_{g}$ states are mixed $J_{S}$ may well be  FM and  small. {Using the energies of the  AFD, FM, and AF-G phases in the expressions}
$$E_{\rm D}-E_{\rm F}=(16J_{L}+8J_{S})
\mu^{2}_{\rm Ni_{\rm II}},\   E_{\rm G}-E_{\rm F}=16J_{S}
\mu^{2}_{\rm Ni_{\rm II}},$$
{we extract  $J_{S}$=--4.1 meV and a small FM $J_{S}$=0.9 meV. 
The energy difference $E_{\rm A}$--$E_{\rm dF}$=8$J_{\perp}$$\mu^{2}_{\rm Ni_{\rm II}}$ of the AF-A (LAO-separated AF-stacked FM LNO planes) and double-FM (the same stacked FM) phases yields a tiny $J_{\perp}$$\simeq$--0.05 meV, as expected due to suppressed hopping through LAO's Al $p$ states very far from the Fermi energy.  The large coupling anisotropy $\alpha$=$J_{L}$/$J_{\perp}$$\simeq$100 and the fact that $J_{S}$  does not contribute  to the magnetic energy  of the AFD phase (see Fig.\ref{figmag}) suggest  using the Ne\'el temperature T$_{\rm N}$=4$\pi$$J_{L}$$S^{2}\!$/ln($\alpha$$\pi^{2}$) of  the 3D anisotropic AF Heisenberg model \cite{anising} as estimate  of the critical temperature: we find T$_{\rm N}$$\simeq$50 K, in good agreement  with 40 K experimentally. (As $\alpha$ only affects T$_{\rm N}$  logarithmically, its exact value is not essential as long as it is large.) 
This  interpretation of the magnetic transition as spin order--disorder  implicitly  assumes}  that  disordered spins fluctuate rapidly enough (say, frequency$\geq$1 MHz) above T$_{\rm N}$  so as not to be revealed by $\mu$SR.

We note in passing that FM and AF-G are also insulating and have  structure, charge, and magnetic pattern largely similar to AFD; the main difference is that in the  FM,  $\mu_{\rm Ni_{\rm IV}}$=0.07 $\mu_{B}$, which we can neglect  compared to $\mu_{\rm Ni_{\rm II}}$ for our present estimate.

\begin{figure}[h]
\includegraphics[width=8.4cm]{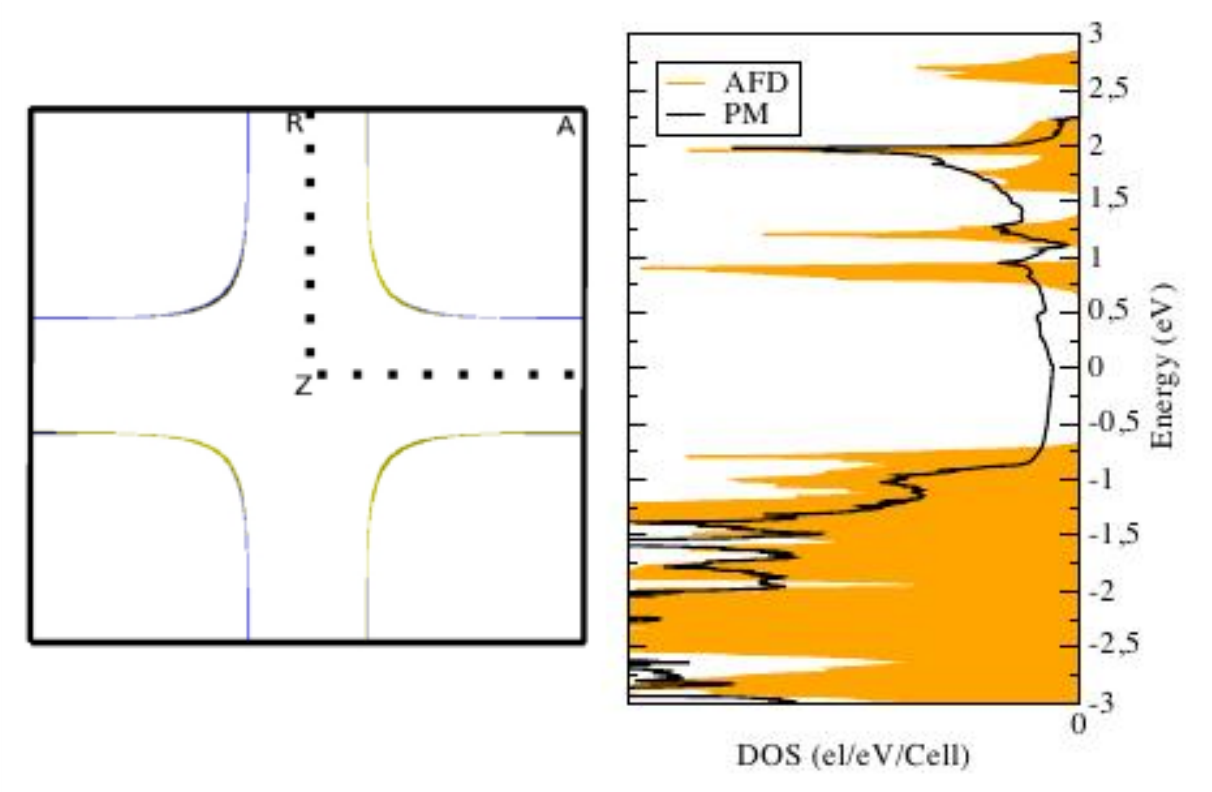}
\caption{(Color on line) Right panel: Total   density of states of the AFD (orange, shaded) and Pauli-PM (solid line)  states (zero is the Fermi energy). Left panel: Fermi surface of the Pauli-PM phase  in the 1$\times$1 Brillouin zone.}
\label{figele}
\end{figure}

\paragraph{Electronic structure and metal-insulator transition$\!\!\!\!$}-- We now address  electronic properties and discuss  mechanisms of  metal-insulator transition. Fig.\ref{figele} reports the total density of  states (DOS), of AFD and Pauli-PM, right panel, and the Fermi surface of the latter (see below), left panel.  The concurrent structural  dimerization and magnetic superstructure open a 1.3-eV indirect electronic gap. The key point is that 
the octahedra distortion is essential to obtain a gap: all undistorted phases are metallic and show no charge transfer. Further, only the Pauli-PM metal is stable among these, while the AF or FM states dimerize spontaneously. 
This suggests that the  transition be  associated to the structural dimerization, and that the transition temperature T$_{\rm MI}$  be identified as that at which the structure un-dimerizes thermally, with attendant gap closure. Using a {stripped-down version of Vineyard's transition-state theory}, we describe the initially full population N$_{0}$$\equiv$N($t$=0) of distorted structural units (contracted and expanded Ni-octahedron pairs) as  undergoing thermal activation out of the low-T ground state. The population N($t$)=N$_{0}$\,$\exp{(-Rt)}$ is  abruptly depleted, i.e. the system removes the distortion and hence the insulating character,   for a sufficient Arrhenius activation rate $R$=$\nu_{0}\,$exp$(-\Delta E/k_{B}{\rm T})$. Since the  Pauli-PM is the only stable undimerized state, we envisage an AFD-PM transition, and  therefore use the AFD-PM energy difference $\Delta$$E$=0.40 eV per Ni pair. With a plausible effective vibrational  prefactor $\nu_{0}$=5 THz \cite{viblanio}, 
an activation rate R between 10$^{-6}$ and 1 Hz (i.e., lifetimes between 280 hrs and 1 sec) corresponds to 
$${\rm T}_{\rm MI}=-\Delta E/[k_{B}(\ln R - \ln \nu_{0})]=110\div150\ {\rm K}
$$
in good agreement with 110 K experimentally \cite{boris}. In closing this section we note that the Ni$_{\rm IV}$-Ni$_{\rm II}$ charge transfer is associated in optical experiments to a spectral weight depletion below 0.4 eV,  identified as a ``charge gap''  \cite{boris}; our best shot at it is the (indeed somewhat larger) electronic gap,   originating from the combined  structure, charge and magnetic ordering. 

{We note that  in the AFD state the valence top states mostly  project on polarized Ni$_{\rm III}$, while conduction states do so on unpolarized Ni$_{\rm IV}$. This ``site-discriminated'' gap opening is quite consistent with the ``site selective Mott transition'' proposed by DMFT calculations \cite{millis}, further suggesting that our method can produce, in specific instances, predictions matching those of sophisticated many-body methods.}

\paragraph*{Fermi surface$\!\!\!\!$}--  Confirming earlier theoretical suggestions \cite{kha,held},  the  Pauli-PM phase has a single-sheet hole-like Fermi surface (Fig.\ref{figele}, left panel) centered at the  1$\times$1 Brillouin zone corner and analogous to optimally-doped cuprates, The  states character  is, however, mixed $e_{g}$ rather than pure $x^{2}$--$y^{2}$, as also found by recent   dynamical mean field calculations \cite{millis2}. 
%
%
The nearly two-dimensional pockets in Fig. \ref{figele} should give rise to quantum oscillation as function of inverse magnetic field, with potentially observable frequencies of about 20 kTesla (compare the 30 kTesla in  e.g. metallic In \cite{dhvin}).

\paragraph{Transport$\!\!\!\!$}-- Since we cannot calculate the T dependence of dielectric response measured in \cite{boris}, we  use  Bloch-Boltzmann theory to calculate the $d$$c$ conductivity, with the goal of  associating  the transition T with the zero of d$\sigma$/dT as suggested in \cite{boris}. We tuned the energy dependence of the relaxation-time  model \cite{bbt}  to reproduce the T-dependence (not the value) of $\sigma$ in the metal phase
using the  Pauli-PM bands. The model was then fed the AFD ground-state bands to obtain its conductivity  vs T.

 \begin{figure}[ht]
\includegraphics[width=8.5 cm]{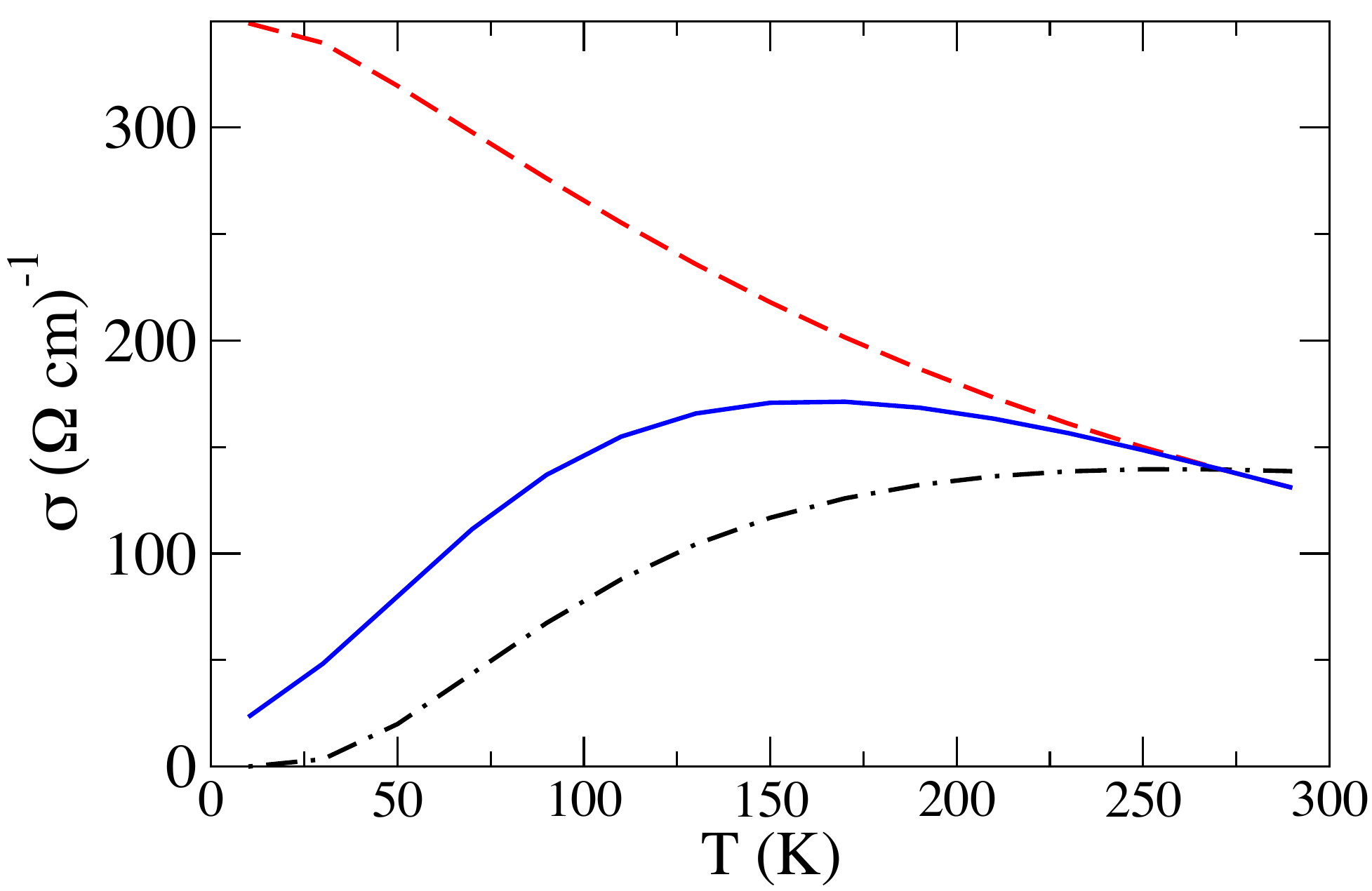}
\caption{Calculated conductivity vs. T for the metallic PM phase (dashed) and $n$-doped AFD phase (dash-dot); solid line: linear interpolation of the metal and insulator curves. The value of $\mu$ is fixed at the vertical dashed line in Fig.\ref{sigma_mu}.  }
\label{sigma_temp}
\end{figure}

We conjecture  that the un-dimerization metal-insulator transition  will cause the conductivity to cross over smoothly from the insulator to the metal.  This  can only be assessed  qualitatively in the present context. First, our method cannot describe the  ``dirty'' metal phase, which  exhibits an unusually  low experimental conductivity; to account for this, we rescale the calculated metal $\sigma$ by a factor 1/15, the ratio of the relaxation time  \cite{boris} for the SL metal phase to that of a normal metal (Al). Second, the insulating phase's  experimental conductivity is much higher than that of our undoped insulator at the relevant temperatures. We assume that this is due to a background impurity of unidentified origin, and thus calculate $\sigma$ in the insulating phase for the chemical potential $\mu$ set to $n$-type.  In Fig.\ref{sigma_temp} we  interpolate linearly the two  $\sigma$'s just discussed (dashed and dash-dotted  lines) vs.T, obtaining a result (solid line) qualitatively   similar  to experiment (\cite{boris}, Fig. S8B of supporting material) and with d$\sigma$/dT=0  around 150 K, fairly consistent with  un-dimerization.  We note in passing that log $\sigma$ in the insulator phase is linear in 1/T as in \cite{boris}. Hopping behavior\cite{mottlno} may be due to local disorder \cite{boris2}, which we cannot assess.

\begin{figure}[h]
\includegraphics[width=8 cm]{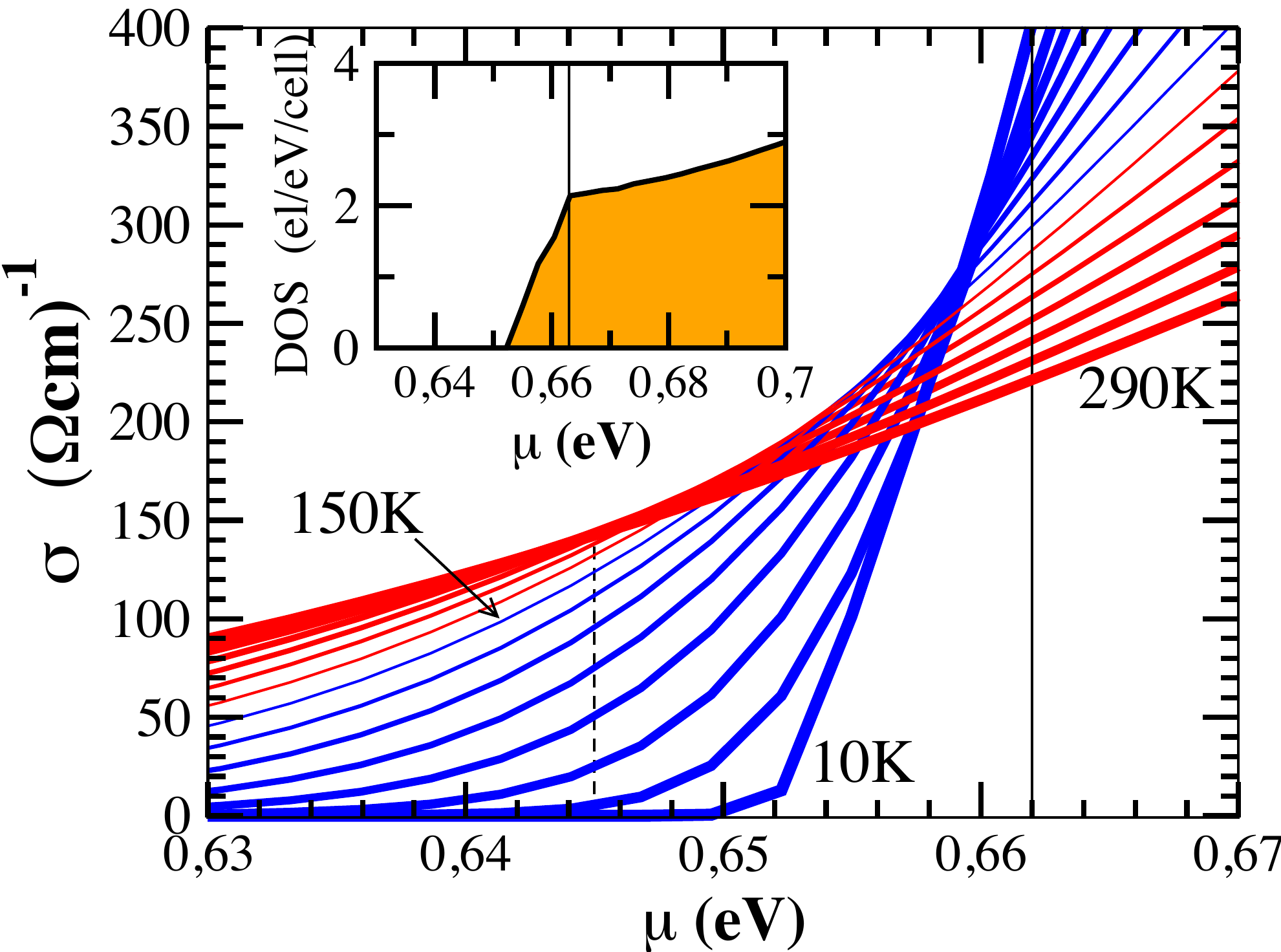}
\caption{ (Color on line) Conductivity  vs. chemical potential and T from 10 K, thickest blue (black) line to 290 K, thickest red (gray) line (20-K steps). Inset: DOS near  conduction edge. Dashed (solid) vertical line marks intermediate regime (metallic crossover).  Chemical potential zero at midgap.}
\label{sigma_mu}
\end{figure}

Conductivity calculations as a function of $\mu$, and specifically in $n$-type conditions, suggest a further possible, concurrent transition mechanism.  $\sigma$($\mu$,T)  in Fig.\ref{sigma_mu} exhibits three distinct
regimes as a function of increasing  $\mu$. Initially, $\sigma$ is insulator-like 
vs.$\mu$ and T,  saturating at high T for fixed $\mu$; then, for $\mu$ just below the conduction edge, $\sigma$ is  insulator-like at low T, and crosses over  at higher T to a normal-metal-like linear decrease; finally, at larger $\mu$,  $\sigma$ decreases with T (and grows with $\mu$) linearly as in a normal metal. This behavior is related to  the slope change of the AFD near-conduction DOS  (inset of Fig.\ref{sigma_mu}), with the crossover to metallic conduction occurring for $\mu$ above the DOS cusp (vertical solid line).
The insulator $\sigma $(T) in Fig.\ref{sigma_temp} is obtained for $\mu$ just below the conduction edge (dashed vertical line in Fig.\ref{sigma_mu}), and has 
 d$\sigma$/dT=0  at about 250 K. This calculation thus shows that an insulator-metal crossover can also be legitimately associated to a conduction-band-edge  Fermi level pinning   (e.g. by shallow-donor defects). Clearly, this mechanism would be preempted by, or at most concurrent with, the un-dimerization transition discussed earlier, which is  robustly rooted  in the structural and magnetic energetics.

\paragraph{Summary$\!\!\!\!$}-- LNO/LAO ultra-short-period SLs have a magnetic charge-ordered insulator ground state, making magnetic and metal-insulator  transitions  at temperatures we estimate in 50-70 K and 110-150 K respectively  to a metal phase with a cuprate-like Fermi surface. This interpretation is in good agreement with available experimental data. While resulting from several cooperative effects, this state is basically  produced by an instability of a "checkerboard" breathing mode (induced in turn, from a chemical viewpoint, by a valency instability). This is indirectly supported by recent experimental evidence of the inverse effect, i.e. the destabilization and metallization of insulating NdNiO$_{3}$ via phonon injection from the LAO substrate \cite{caviglia}. 

We thank A. Boris for helpful discussions. Work supported in part by  EU (project OxIDes), IIT (Seed project NEWDFESCM), MIUR (PRIN 2DEG-FOXI), and Fondazione Banco di Sardegna  grants. Computing resources  provided by CASPUR, CINECA, and Cybersar.


\begin{thebibliography}{99}
\bibitem{notedmft} The PM-phase Fermi surface is similar to that in \cite{held}, and the pseudo-gapped DOS in metallic magnetic phases (not discussed  here) is similar to that of DMFT at large U  \cite{held2}.
\end{thebibliography}
\end{document}